\documentclass{article}

\usepackage{algorithm}
\usepackage{algpseudocode}
\usepackage{tikz}
\usepackage{amsthm}
\usepackage{url}

\newtheorem{theorem}{Theorem}[section]

\newtheorem{proposition}[theorem]{Proposition}

\graphicspath{{./graphics/}}

\title{Pruning Algorithms for Pretropisms \\
of Newton Polytopes\thanks{This material is based upon work 
supported by the National Science Foundation under Grant No. 1440534.}}

\author{Jeff Sommars \and Jan Verschelde}

\date{University of Illinois at Chicago \\
Department of Mathematics, Statistics, and Computer Science \\
851 S. Morgan Street (m/c 249), Chicago, IL 60607-7045, USA \\
{\tt \{sommars1,janv\}@uic.edu}}

\begin{document}
\maketitle

\begin{abstract}
Pretropisms are candidates for the leading exponents of Puiseux series 
that represent
positive dimensional solution sets of polynomial systems. We propose a new
algorithm to both horizontally and vertically prune the tree of edges
of a tuple of Newton polytopes. We provide experimental results with our preliminary implementation in Sage that demonstrates that our algorithm compares favorably to the definitional algorithm.
\end{abstract}

\section{Introduction}

Almost all polynomial systems arising in applications are sparse,
as few monomials appear with nonzero coefficients,
relative to the degree of the polyomials.
Polyhedral methods exploit the sparse structure of a polynomial system.
In the application of polyhedral methods to compute positive dimensional
solution sets of polynomial systems,
we look for series developments of the solutions,
and in particular we look for Puiseux series~\cite{Mau80}.
The leading exponents of Puiseux series are called {\em tropisms}.
The {\em Newton polytope} of a polynomial in several variables
is the convex hull of the exponent tuples of the monomials that
appear with nonzero coefficient in the polynomial.

In~\cite{BJSST07}, polyhedral methods were defined
in tropical algebraic geometry.
We refer to~\cite{MS15} for a textbook introduction
to tropical algebraic geometry.  Our textbook reference
for definitions and terminology of polytopes is~\cite{Zie95}.

Our problem involves the intersection of polyhedral cones.
A {\em normal cone} of a face $F$ of a polytope $P$ is the convex cone 
generated by all of the facet normals of facets which contain $F$. 
The {\em normal fan} of a polytope $P$ is the union of all of the normal
cones of every face of $P$.
Given two fans $F_1$ and $F_2$, their {\em common refinement} 
$F_1 \wedge F_2$ is defined as
\begin{equation} \label{eqdefrefinement1}
   F_1 \wedge F_2 =
   \bigcup_{\begin{array}{c} C_1 \in F_1 \\ C_2 \in F_2 \end{array}}
     C_1 \cap C_2.
\end{equation}
As the common refinement of two fans is again a fan,
the common refinement of three fans $F_1$, $F_2$, and $F_3$
may be computed as $(F_1 \wedge F_2) \wedge F_3$.

\noindent {\bf Problem Statement.}
Given the normal fans $(F_1, F_2, \ldots, F_n)$
of the Newton polytopes $(P_1, P_2, \ldots, P_n)$,
a {\em pretropism} is a ray in a cone $C$,
\begin{equation}
  C = C_1 \cap C_2 \cap \cdots \cap C_n
  \in F_1 \wedge F_2 \wedge \cdots \wedge F_n,
\end{equation}
where each $C_i$ is the normal cone to some $k_i$-dimensional face of $P_i$,
for $k_i \geq 1$, for $i=1, 2, \ldots, n$.
Our problem can thus be stated as follows:
given a tuple of Newton polytopes, compute all pretropisms.

We say that two pretropisms are equivalent if they are both perpendicular
to the same tuples of faces of the Newton polytopes.
Modulo this equivalence, there are only a finite number of pretropisms.
Ours is a difficult problem because of the dimension restrictions
on the cones.  In particular, the number of pretropisms can be very small 
compared to the total number of cones in the common refinement.

Pretropisms are candidates tropisms, but not every pretropism is 
a tropism, as pretropisms depend only on the Newton polytopes of the system.
For polynomial systems with sufficiently generic coefficients,
every tropism is also a pretropism.  See~\cite{BV16} for an example.

\noindent {\bf Related Work.}
A tropical prevariety was introduced in~\cite{BJSST07}
and Gfan~\cite{Jen08} is a software system to compute the common 
refinement of the normal fans of the Newton polytopes.
Gfan relies on the reverse search algorithms~\cite{AF92}
in cddlib~\cite{FP96}.

The problem considered in this paper is a generalization of the 
problem to compute the mixed volume of a tuple of Newton polytopes,
for which pruning methods were first proposed in~\cite{EC95}.
Further developments can be found in~\cite{GL03} and~\cite{MTK07},
with corresponding free software packages MixedVol~\cite{GLW05}
and DEMiCS~\cite{MT08}.  A recent parallel implementation along
with a complexity study appears in~\cite{Mal15}.
The relationship between triangulations and the mixed subdivisions
is explained and nicely illustrated in~\cite{DRS10}.

The main difference between mixed volume computation and the
computation of the tropical prevariety is that in a mixed volume
computation the vertices of the polytopes are lifted randomly,
thus removing all degeneracies.  This lifting gives the powers of
an artificial parameter.  In contrast, in a Puiseux series development
of a space curve, the first variable is typically identified as the
parameter and the powers of the first variable in the given polynomials
cannot be considered as random.

A practical study on various software packages for
exact volume computation of a polytope is described in~\cite{BEF00}.
Exact algorithms on Newton polytopes are discussed in~\cite{EFK12}.
The authors of~\cite{EF14} present an
experimental study of approximate polytope volume computation.
In~\cite{EFG16}, a polynomial-time algorithm is presented to
compute the edge skeleton of a polytope.
Computing integer hulls of convex polytopes can be done
with polymake~\cite{AGHJLPR15}.

\noindent {\bf Our contributions and organization of the paper.}
In this paper we outline two different types of pruning algorithms
for the efficient computation of pretropisms.
We report on a preliminary implementation in Sage~\cite{Sage}
and illustrate the effectiveness on a parallel computer
for various benchmark problems. This paper extends the results of our
EuroCG paper~\cite{SV16} as well as~\cite{SV15}.

\section{Pruning Algorithms}

\subsection{Horizontal and Vertical Pruning Defined}

Because we are interested only in those cones of the common refinement
that contain rays perpendicular to faces of dimension one or higher,
we work with the following modification of~(\ref{eqdefrefinement1}):
\begin{equation} \label{eqdefrefinement2}
   F_1 \wedge_1 F_2 =
   \bigcup_{
      \begin{array}{c}
         C_1 \in F_1, C_1 \perp \mbox{ edge of $P_1$ } \\
         C_2 \in F_2, C_2 \perp \mbox{ edge of $P_2$ }
      \end{array}}
     C_1 \cap C_2.
\end{equation}
The $\wedge_1$ defines the {\em vertical pruning} as the replacement
of $\wedge$ by $\wedge_1$ in $(F_1 \wedge F_2) \wedge F_3$ so we compute
$(F_1 \wedge_1 F_2) \wedge_1 F_3$.
Cones in the refinement that do not satisfy the dimension restrictions
are pruned away in the computations.
Our definition of vertical pruning is currently incomplete, 
but we will refine it in~\ref{pruningrevisited} after we have formally
defined our algorithms.

The other type of pruning, called {\em horizontal pruning} is
already partically implicitly present in 
the $\bigcup$ operator of~(\ref{eqdefrefinement2}),
as in a union of sets of cones, every cone is collected only once,
even as it may originate as the result of many different cone intersections.
With horizontal pruning we remove cones of $F_1 \wedge_1 F_2$
which are contained in larger cones.
Formally, we can define this type of pruning via the $\wedge_2$ operator:
\begin{equation}
   F_1 \wedge_2 F_2 =
   \bigcup_{
      \begin{array}{c}
         C \in F_1 \wedge_1 F_2, C \not\subset C' \\
         C' \in F_1 \wedge_1 F_2 \setminus \{ C \}
      \end{array}}
     C.
\end{equation}

\subsection{Pseudo Code Definitions of the Algorithms}

Algorithm~\ref{horizontalprune} sketches the outline of our algorithm
to compute all pretropisms of a set of $n$ polytopes.
Along the lines of the gift wrapping algorithm,
for every edge of the first polytope we take the plane that contains
this edge and consider where this plane touches the second polytope.
Algorithm~\ref{algedgeskeleton} starts exploring the edge skeleton defined
by the edges connected to the vertices in this touching plane.

\begin{algorithm}[hbt]
\begin{algorithmic}[1]
\caption{Explores the skeleton of edges to find pretropisms
         of a polytope and a cone.}
\label{algedgeskeleton}
\Function{ExploreEdgeSkeleton}{Polytope $P$, Cone $C$}
\State $r$ := a random ray inside $C$
\State ${\rm in}_{r}(P)$ := vertices of $P$ with minimal inner product with $r$
\State EdgesToTest := all edges of $P$ that have vertices in ${\rm in}_{r}(P)$
\State Cones := $\emptyset$
\State TestedEdges := $\emptyset$
\While{EdgesToTest $\not= \emptyset$}
\State $E$ := pop an edge from EdgesToTest
\State $C_E$ := normal cone to $E$
\State ShouldAddCone := False
\If{$C_E$ contains $C$}
\State ConeToAdd := $C$
\State ShouldAddCone := True
\ElsIf
{$C \cap C_E \not= \{ 0 \}$}
\State ConeToAdd := $C \cap C_E$
\State ShouldAddCone := True
\EndIf

\If{ShouldAddCone}
\State Cones := Cones $\cup~$ConeToAdd
\State Edges := Edges $\cup~E$
\For{each neighboring edge $e$ of $E$}
\If{$e \not\in$ TestedEdges}
\State EdgesToTest := EdgesToTest$\cup e$
\EndIf
\EndFor
\EndIf
\State TestedEdges := TestedEdges $\cup~E$
\EndWhile
\State \Return Cones
\EndFunction
\end{algorithmic}
\end{algorithm}

The exploration of the neighboring edges corresponds to tilting
the ray~$r$ in Algorithm~\ref{algedgeskeleton}, as in rotating a 
hyperplane in the gift wrapping method.
One may wonder why the exploration of the edge skeleton in 
Algorithm~\ref{algedgeskeleton} needs to continue after 
the statement on line~4. This is because the cone $C$ has
the potential to intersect many cones in $P$, particularly if $P$ has
small cones and $C$ is large. Furthermore it is reasonable to wonder why we 
bother checking cone containment when computing the intersection
of two cones provides more useful information. Checking cone
containment means checking if each of the generators of $C$ is contained
in $C_E$, which is a far less computationally expensive operation
than computing the intersection of two cones.

In the Newton-Puiseux algorithm to compute series expansions, we are 
interested only in the edges on the lower hull of the Newton polytope,
i.e. those edges that have an upward pointing inner normal.
For Puiseux for space curves, the expansions are normalized so that
the first exponent in the tropism is positive.
Algorithm~\ref{horizontalprune} is then easily adjusted so that
calls to the edge skeleton computation of Algorithm~\ref{algedgeskeleton}
are made with rays that have a first component that is positive.

\begin{algorithm}[hbt]
\caption{Finds pretropisms for a given set of polytopes}
\label{horizontalprune}
\begin{algorithmic}[1]
\Function{FindPretropisms}{Polytope $P_1$, Polytope
   $P_2, \ldots,$ Polytope $P_{n}$}

\State Cones := set of normal cones to edges in $P_1$
\For{i := 2 to n}
\State NewCones := $\emptyset$
\For{Cone in Cones}
\State NewCones := NewCones $\cup$ ExploreEdgeSkeleton($P_i$, Cone)
\EndFor
\For{Cone in NewCones}
\If{Cone is contained within another cone in NewCones}
\State NewCones := NewCones - Cone
\EndIf
\EndFor
\State Cones := NewCones
\EndFor
\State Pretropisms := set of generating rays for each cone in Cones
\State \Return Pretropisms
\EndFunction
\end{algorithmic}
\end{algorithm}

\subsection{Correctness}

To see that these algorithms will do what they claim,
we must define an additional term. A {\em pretropism graph} is
the set of edges for a polytope that have normal cones
intersecting a given cone. We will now justify why 
the cones output by Algorithm~\ref{algedgeskeleton} correspond
to the full set of cones that live on a pretropism graph.

\begin{theorem} \label{theorem1}
Pretropism graphs are connected graphs.
\end{theorem}
\noindent {\em Proof.}
Let $C$ be a cone, and let $P$ be a polytope with edges $e_1, e_2$ such that
they are in the pretropism graph of $C$.
Let $C_1$ be the cone of the intersection of the normal cone of
$e_1$ with $C$, and let $C_2$ be the cone of the intersection 
of the normal cone of $e_2$ and $C$.
If we can show that there exists a path between $e_1$ and $e_2$
that remains in the pretropism graph, then the result will follow.

Let $n_1$ be a normal to $e_1$ that is also in $C_1$ 
and let $n_2$ be a normal to $e_2$ that is also in $C_2$.
Set $n = tn_1 + (1-t)n_2$ where $0 \le t \le 1$. 
Consider varying $t$ from 0 to 1; 
this creates the cone $C_n$, a cone which must 
lie within $C$, as both $n_1$ and $n_2$ lie in that cone. 
As $n$ moves from 0 to 1, it will progressively intersect new faces of $P$
that have all of their edges in the pretropism graph. Eventually, this 
process terminates when we reach $e_2$, and we have constructed
a path from $e_1$ to $e_2$. Since a path
always exists, we can conclude that pretropism graphs are 
connected graphs.
~\qed

Since pretropism graphs are connected, Algorithm~\ref{algedgeskeleton}
will find all cones of edges on the pretropism graph. In 
Algorithm~\ref{horizontalprune}, we iteratively explore the edge skeleton
of polytope $P_i$, and use the pruned set of cones to explore $P_{i+1}$. From
this, it is clear that Algorithm~\ref{horizontalprune} will compute
the full set of pretropisms.

\subsection{Analysis of Computational Complexity}

In estimating the cost of our algorithm to compute all pretropisms,
we will first consider the case when there are two polytopes.
We will take the primitive operation of computing pretropisms to be
the number of cone intersections performed, as that number will
drive the time required for the algorithm to complete.
For a polytope $P$, denote by $n_e(P)$ its number of edges.
The upper bound on the number of primitive operations for two
polytopes $P_1$ and $P_2$ is the product $n_e(P_1) \times n_e(P_2)$,
while the lower bound equals the number of pretropisms.

Denote by $E_{P,e}$ the pretropism graph 
resting on polytope~$P$ corresponding to the ray determined by edge~$e$.
Let $n_e(E_{P,e})$ denote the number of edges in~$E_{P,e}$.

\begin{proposition}
The number of primitive operations in Algorithm~\ref{horizontalprune}
on two polytopes $P_1$ and $P_2$ is bounded by
\begin{equation} \label{eqcostbound}
   \sum_{i=1}^{n_e(P_1)} n_e(E_{P_2,e_i}),
\end{equation}
where $e_i$ is the $i$-th edge of~$P_1$.
\end{proposition}

As $E_{P,e}$ is a subset of the edges of~$P$:
$n_e(E_{P,e}) \leq n_e(P)$.
Therefore, the bound in~(\ref{eqcostbound}) is smaller
than $n_e(P_1) \times n_e(P_2)$.

To interpret~(\ref{eqcostbound}), recall that Algorithm~\ref{horizontalprune}
takes a ray from inside a normal cone to an edge of the first polytope
for the exploration of the edge graph of the second polytope.
If we take a simplified view on the second polytopes as a ball,
then shining a ray on that ball will illuminate at most half of
its surface.  
If we use the estimate: $n_e(E_{P_2, e_i}) \approx n_e(P_2)/2$,
then Algorithm~\ref{horizontalprune} cuts the the upper bound on the 
number of primitive operations in half.

Estimating the cost of the case of $n$ polytopes follows naturally
from the cost analysis of the case of 2 polytopes.
For $n$ polytopes, the upper bound
on the number of primitive operations required is the product
$n_e(P_1) \times n_e(P_2) \times  \ldots \times n_e(P_{n})$.
\begin{proposition}
The number of primitive operations in Algorithm~\ref{horizontalprune}
on $n$ polytopes $P_1$, $P_2, \ldots, P_{n}$ is bounded by
\begin{equation} \label{eqncostbound}
   \sum_{i=1}^{n_e(P_1)} \left( \prod_{j=2}^{n} n_e(E_{P_j,e_i}) \right)
\end{equation}
where $e_i$ is the $i$-th edge of~$P_1$.
\end{proposition}
Again, if we use the estimate that $n_e(E_{P_j, e_i}) \approx n_e(P_j)/2$,
then Algorithm~\ref{horizontalprune} reduces the upper bound on the number of
primitive operations by $\frac{1}{2^{n-1}}$. This estimate depends entirely
on the intuition that we are cutting the number of comparisons in half. This
estimate may not hold in the case when we have large lineality spaces, and
thus have huge input cones. This situation can be partially remedied by
sorting the input polytopes from smallest dimension of lineality space
to highest. This seeds Algorithm~\ref{horizontalprune} with the smallest
possible input cones.

\subsection{Horizontal and Vertical Pruning Revisited}
\label{pruningrevisited}

The definitional algorithm of pretropism can be interpreted as creating
a tree structure. From a root node, connect the cones of $P_1$. On the 
next level of the tree, place the cones resulting from intersecting each
of the cones of $P_2$ with the cones of $P_1$, connecting the new cones
with the cone from $P_1$ that they intersected. It is likely that
at this level of the tree there are many cones that are empty. Continue
this process creating new levels of cones representing the intersection
of the previous level of cones with the next polytope until the $n$th 
polytope has been completed. The cones at the $n$th layer of the tree
represent the cones generated by pretropisms.

\begin{figure}[h]
\begin{center}
\begin{picture}(200,240)(0,0)
\put(0,125){
\begin{tikzpicture}
\node (root) at (3,6) [rectangle] {};
\node (A) at (1.5,4.5)  [rectangle]{A};
\node (B) at (3, 4.5) [rectangle]{B};
\node (C) at (4.5,4.5)  [rectangle]{C};
\draw (root) edge (A)  (root) edge (B) (root) edge (C);

\node (D) at (0,3.5) {D};
\node (E) at (0.5,3.5) {E};
\node (F) at (1,3.5) {F};
\node (G) at (1.5,3.5) {G};
\draw (A) edge (D)  (A) edge (E)  (A) edge (F)  (A) edge (G);

\node (H) at (2.5,3.5) {H};
\node (I) at (3,3.5) {I};
\node (FF) at (3.5,3.5) {F};
\draw (B) edge (H)  (B) edge (I)  (B) edge (FF);

\node (J) at (4.5,3.5) {J};
\node (HH) at (5,3.5) {H};
\node (GG) at (5.5,3.5) {G};
\draw (C) edge (J)  (C) edge (HH) (C) edge (GG);

\node (K) at  (-0.5,2.5) {K};
\node (L) at  (0,2.5) {L};
\draw (D) edge (K) (D) edge (L);

\node (M) at  (0.4,2.5) {M};
\node (N) at  (0.7,2.5) {N};
\node (O) at  (1.0,2.5) {O};
\node (P) at  (1.3,2.5) {P};
\draw (F) edge (M) (F) edge (N) (F) edge (O) (F) edge (P);

\node (Q) at (1.7,2.5) {Q};
\draw (G) edge (Q);

\node (R) at  (2.2,2.5) {R};
\node (S) at  (2.5,2.5) {S};
\draw (H) edge (R) (H) edge (S);

\node (T) at  (2.8,2.5) {T};
\node (U) at  (3.1,2.5) {U};
\draw (I) edge (T) (I) edge (U);

\node (MM) at  (3.4,2.5) {M};
\node (NN) at  (3.7,2.5) {N};
\node (OO) at  (4.0,2.5) {O};
\node (PP) at  (4.3,2.5) {P};
\draw (FF) edge (MM) (FF) edge (NN) (FF) edge (OO) (FF) edge (PP);

\node (RR) at  (4.9,2.5) {R};
\node (SS) at  (5.3,2.5) {S};
\draw (HH) edge (RR) (HH) edge (SS);

\node (QQ) at (5.8,2.5) {Q};
\draw (GG) edge (QQ);
\end{tikzpicture}
}
\put(0,0){
\begin{tikzpicture}
\node (root) at (3,6) [rectangle] {};
\node (A) at (1.5,4.5)  [rectangle]{A};
\node (B) at (3, 4.5) [rectangle]{B};
\node (C) at (4.5,4.5)  [rectangle]{C};
\draw (root) edge (A)  (root) edge (B) (root) edge (C);

\node (D) at (0,3.5) {D};
\node (E) at (0.5,3.5) {E};
\node (F) at (1,3.5) {F};
\node (G) at (1.5,3.5) {G};
\draw (A) edge (D)  (A) edge (E)  (A) edge (F)  (A) edge (G);

\node (H) at (2.5,3.5) {H};
\node (I) at (3,3.5) {I};
\node (FF) at (3.5,3.5) {F};
\draw (B) edge (H)  (B) edge (I)  (B) edge (FF);

\node (J) at (4.5,3.5) {J};
\node (HH) at (5,3.5) {H};
\node (GG) at (5.5,3.5) {G};
\draw (C) edge (J)  (C) edge (HH) (C) edge (GG);

\node (K) at  (-0.5,2.5) {K};
\node (L) at  (0,2.5) {L};
\draw (D) edge (K) (D) edge (L);

\node (M) at  (0.4,2.5) {M};
\node (N) at  (0.7,2.5) {N};
\node (O) at  (1.0,2.5) {O};
\node (P) at  (1.3,2.5) {P};
\draw (F) edge (M) (F) edge (N) (F) edge (O) (F) edge (P);

\node (Q) at (1.7,2.5) {Q};
\draw (G) edge (Q);

\node (R) at  (2.2,2.5) {R};
\node (S) at  (2.5,2.5) {S};
\draw (H) edge (R) (H) edge (S);

\node (T) at  (2.8,2.5) {T};
\node (U) at  (3.1,2.5) {U};
\draw (I) edge (T) (I) edge (U);

\end{tikzpicture}
}
\end{picture}
\caption{Nodes A, B, C represent cones to $P_1$.
Nodes D, E, F, and G represent intersections of cone A 
with cones to~$P_2$, etc.
Nodes K and L represent intersections of cone D with cones to~$P_3$, etc.
Duplicate nodes are removed from the second tree at the bottom. }
\label{figconegraphs}
\end{center}
\end{figure}
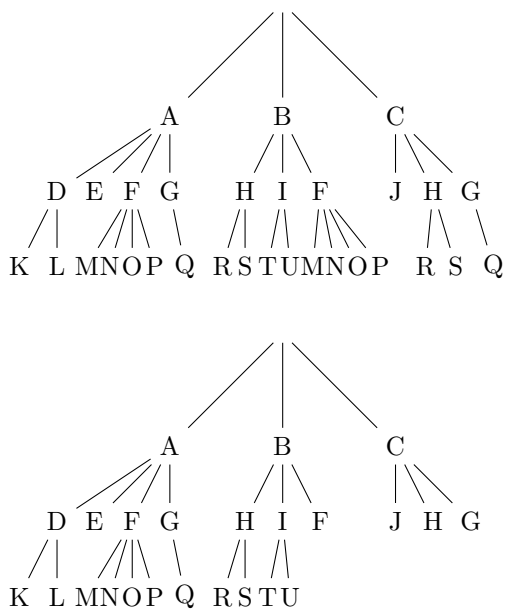

Our algorithms can be seen to improve on this basic tree structure in
two distinct ways. Algorithm~\ref{algedgeskeleton} reduces the number of
comparisons needed through exploring the edge skeletons of the polytopes.
Because of this, there are many times that we do not perform cone
intersections that will result in 0 dimensional cones. 
From the perspective of the tree, this is akin to avoiding drawing edges 
to many 0 dimensional cones; we call this vertically pruning the tree. 
We horizontally prune the tree through Algorithm~\ref{horizontalprune}
which reduces the number of cones necessary to follow for a given level.
This is illustrated in Figure~\ref{figconegraphs}.
By both horizontally and vertically pruning the tree of cones, 
we are able to avoid performing many unnecessary cone intersections. 
We will demonstrate the benefits of pruning experimentally in the
sections to come.

\section{Implementation}
Our algorithm takes as input a modified version of the data structure output
by the gift wrapping algorithm to compute convex hulls. It conceptually exploits
the connectivity between vertices, edges, and facets, but
only requires the edge skeleton of the polytope. To accomplish
this, we created edge objects that had vertices, references to their
neighboring edges, and the normal cone of the edge. When polytopes
were not full dimensional, we included the generating rays of the lineality
space when we created the normal cones. This has the negative effect of
increasing the size of the cone, but is essential for the algorithm to work.

\subsection{Code}
We developed a high level version of Algorithm~\ref{horizontalprune}
in Sage~\cite{Sage}, using its modules for lattice polytopes~\cite{Nov11}, 
and polyhedral cones~\cite{BH11}.
To compute the intersections of cones, Sage uses PPL~\cite{BHZ08}.
Our preliminary code is available
at https://github.com/sommars/GiftWrap.

\subsection{Parallelism}
To improve the performance of our core algorithms, we also
implemented high level parallelism. We used the built in
Python queue structure to create a job queue of cones
in lines 5 through 7 of Algorithm~\ref{horizontalprune}. Each call to
the Algorithm~\ref{algedgeskeleton} is done independently
on a distinct process, using the computers resources more
efficiently.

Additionally, we have implemented parallelism in checking the cone
containments in lines 8 through 12 of Algorithm~\ref{algedgeskeleton}.
To check if cone $C_1$ is contained within another cone $C_2$, it requires
checking if each of the linear equations and inequalities of $C_1$ is or is
not restricted by each of the linear equations and inequalities of $C_2$.
However, there is much overlap between cones, with many distinct
cones sharing some of the same linear equations or inequalities. Because
of this, we optimized by creating a lookup table to avoid performing
duplicate calculations. However, with large benchmark problems, creation
of this table becomes prohibitively slow. To amend this problem,
we parallelized the creation of the table, with distinct processes
performing distinct calculations.


\section{Generic Experiments}

To test the algorithms, we generate $n-1$ simplices spanned by integer
points with coordinates uniformly generated within the range of 0 to 30.
This input corresponds to considering systems of $n-1$ sparse Laurent 
polynomials in $n$ variables with $n+1$ monomials per equation.
We can compare with the mixed volume computation if we add one extra 
linear equation to the Laurent polynomial system.
Then the mixed volume of the $n$-tuple will give the sum of the degrees
of all the curves represented by the Puiseux series.
Assuming generic choices for the coefficients, the degrees of the curves
can be computed directly from the tropisms, as used in~\cite{Ver09} and
applied in~\cite{AV12,AV13}.

Denoting by ${\rm MV}(P)$ the mixed volume of an $n$-tuple $P$
of Newton polytopes:
\begin{equation}
   {\rm MV}(P) = \sum_{{\bf v}}
    \left( \max_{i=1}^n v_i - \min\left( \min_{i=1}^n v_i, 0 \right) \right)
\end{equation}
where the sum ranges over all tropisms~{\bf v}.

All computations were done on a
2.6 GHz Intel Xeon E5-2670 processor in a Red Hat Linux workstation 
with 128 GB RAM using 32 threads.
When performing generic tests, the program
did not perform any cone containment tests because no cone can be
contained in another cone in this case.

\subsection{Benchmarking}

Table~\ref{tabdata} shows a comparison between the two distinct
methods of computing pretropisms. The mixed volume was
computed with the version of MixedVol~\cite{GLW05},
available in PHCpack~\cite{Ver99} since version 2.3.13.
For systems with generic coefficients, the mixed volume equals the
number of isolated solutions~\cite{Ber75}.
While a fast multicore workstation
can compute millions of solutions, a true supercomputer will be needed
in the case of billions of solutions.
For larger dimensions, the new pruning method dominates the
method suggested by the definition of pretropism.

\begin{table}[ht]
\begin{center}
\caption{Comparisons between the definitional and our pruning method,
for randomly generated generic simplices.  Timings are listed in seconds.}
  \begin{tabular}{|c||r|r||r|r|} \hline
    $n$ & Definitional & Pruning & \#Pretropisms & Mixed Volume \\
 \hline \hline
     3  & 0.008 &     0.20 &     7~~~ &           319 \\ \hline
     4  & 0.11 &     0.42 &    18~~~ &         7,384 \\ \hline
     5  & 1.33 &     0.76 &    58~~~ &       152,054 \\ \hline
     6  & 13.03 &    2.75 &   171~~~ &     4,305,758 \\ \hline
     7  & 243.88 &   20.17 &   614~~~ &    91,381,325 \\ \hline
     8  & 2054.11 & 220.14 & 1,878~~~ & 2,097,221,068 \\ \hline
  \end{tabular}
\end{center}
\label{tabdata}
\end{table}

\subsection{Number of Cone Intersections}

Another way that the definitional algorithm can be compared to our
new algorithm is through comparing the number of cone intersections
required for each algorithm. Table~\ref{tabconecomparisongeneric} contains a comparison
of these numbers. A large number of trials were performed at each dimension so we could
conclude statistically if our mean number of intersections differed from the 
number of intersections required by the cone intersection algorithm. To test
this hypothesis, we performed $t$-tests using the statistical software package R~\cite{R}.
For every dimension from 3 to 8, we were able to reject the null hypothesis
that they had the same mean and we were able to conclude that the new
algorithm has a lower mean number of intersections ($p< 2\times10^{-16}$
for every test). We had estimated the cost to be an improvement by a
factor of $\frac{1}{2^{n-1}}$, but experimentally we found a greater
improvement as can be seen in Table~\ref{tabconecomparisongeneric}.

\begin{table}[ht]
\begin{center}
\caption{Average number of cone intersections required for each algorithm,
comparing the definitional algorithm with our pruning algorithm for generic inputs.
The second to last column contains the ratio
predicted by our cost estimate and the final column contains the actual ratio.}
  \begin{tabular}{|r||r|r||r|r|} \hline
    $n$ & Definitional &  Pruning & Predicted Ratio & Actual Ratio\\
 \hline \hline
     3  &     36 &     29 & 0.5 & 0.72 \\ \hline
     4  &     1,000 &      288 & 0.25 & 0.288 \\ \hline
     5  &    50,625 &   2,424  & 0.125 & 0.0478 \\ \hline
     6  &   4,084,101 &   18,479 & 0.0625 & 0.00452 \\ \hline
     7  &   481,890,304 & 145,134 & 0.03125 & 0.000301  \\ \hline
     8  & 78,364,164,096 &  1,150,386 & 0.015625 & 0.0000147  \\ \hline
  \end{tabular}
\label{tabconecomparisongeneric}
\end{center}
\end{table}

\subsection{Comparison with Gfan}

In the generic case, our code is competitive with Gfan.
Table~\ref{tabgfancomparison} contains timing comparisons, with
input polynomials determined as they were previously determined; 
the timings in the Gfan column were obtained by running the current
version 0.5 of Gfan~\cite{gfan}.

\begin{table}[ht]
\begin{center}
\caption{Comparisons between Gfan and our implementation,
for dimensions 3 through 8.  Timings are listed in seconds.}
  \begin{tabular}{|r||r|r|} \hline
    $n$ &   Gfan  &  Pruning \\
 \hline \hline
     3  &    0.036 &  0.12 \\ \hline
     4  &     0.23 &  0.25 \\ \hline
     5  &     2.03 &   0.80 \\ \hline
     6  &    23.49 &  10.73 \\ \hline
     7  &   299.32 &  49.53 \\ \hline
     8  & 3,764.83 & 540.32  \\ \hline
  \end{tabular}
\label{tabgfancomparison}
\end{center}
\end{table}

\section{Benchmark Polynomial Systems}
Many of the classic mixed volume benchmark problems
like Katsura-$n$, Chandra-$n$, eco-$n$, and
Noonberg-$n$ are inappropriate to use as benchmark systems for 
computing pretropisms.
A good testing system needs to have a positive dimensional solution set
as well as being a system that can be scaled up in size.
The aforementioned mixed volume benchmark problems all lack
positive dimensional solution sets, so we did not perform tests on them.
We have found the cyclic $n$-roots problem
to be the most interesting system that fulfills
both criteria, as there are a variety of sizes of solution sets
within them and the difficulty of computing
pretropisms increases slowly. We also provide
experimental data for the $n$-vortex and the $n$-body
problem, but these problems quickly become uncomputable
with our prototype Sage implementation.

\subsection{Cyclic-$n$ Experiments}

The cyclic $n$-roots problem asks for the solutions
of a polynomial system, commonly formulated as

\begin{equation} \label{eqcyclicsys}
 \left\{
   \begin{array}{c}
     x_{0}+x_{1}+ \cdots +x_{n-1}=0 \\
     i = 2, 3, 4, \ldots, n-1: 
      \displaystyle\sum_{j=0}^{n-1} ~ \prod_{k=j}^{j+i-1}
      x_{k~{\rm mod}~n}=0 \\
     x_{0}x_{1}x_{2} \cdots x_{n-1} - 1 = 0. \\
    \end{array}
 \right.
\end{equation}

This problem is important in the study of biunimodular vectors,
a notion that traces back to Gauss, as stated in~\cite{FR15}.
In~\cite{Bac89}, Backelin showed that if $n$ has a divisor that
is a square, i.e. if $d^2$ divides $n$ for $d \geq 2$, then
there are infinitely many cyclic $n$-roots.
The conjecture of Bj{\"{o}}rck 
and Saffari~\cite{BS95}, \cite[Conjecture~1.1]{FR15}
is that if $n$ is not divisible by a square, 
then the set of cyclic $n$-roots is finite. As shown in~\cite{AV12},
the result of Backelin can be recovered by polyhedral methods.

Instead of directly calculating the pretropisms of the Newton polytopes
of the cyclic $n$-root problem,
we chose to calculate pretropisms of the reduced cyclic $n$-root problem. 
This reformulation~\cite{Emi94} is obtained by performing the substitution 
$x_i = \frac{y_i}{y_0}$ for $i = 0 \dots n-1$. 
Clearing the denominator of each equation leaves the first $n-1$ equations 
as polynomials in $y_1, \dots y_{n-1}$. 
We compute pretropisms of the Newton polytopes
of these $n-1$ equations because they yield meaningful sets of pretropisms.
Calculating with the reduced cyclic $n$-roots problem 
has the benefit of removing
much of the symmetry present in the standard cyclic $n$-roots problem, 
as well
as decreasing the ambient dimension by one. 
Unlike the standard cyclic $n$-roots problem,
some of the polytopes of the reduced cyclic $n$-roots problem
are full dimensional,
which leads to calculation speed ups. A simple transformation can be
performed on the pretropisms we calculate of reduced cyclic $n$-root problem 
to convert them to the pretropisms of cyclic $n$-root problem, so calculating
the pretropisms of reduced cyclic $n$-roots problem is equivalent to 
calculating the pretropisms of the cyclic $n$-roots problem.

Table~\ref{tabdatacyclic} shows how our implementation scales
with time. As with the generic case, our implementation
shows great improvement over the definitional algorithm
as $n$ becomes larger. For $n > 8$, the definitional
algorithm was too inefficient to terminate in the time
allotted.

\begin{table}[hbt]
\begin{center}
\caption{Comparisons between the definitional and our pruning method for 
reduced cyclic-$n$.  Timings are listed in seconds.}
  \begin{tabular}{|c||r|r||r|r|} \hline
    $n$ & Definitional & Pruning & \#Pretropisms & Mixed Volume \\
 \hline \hline
     4  & 0.02 & 0.62 & 2 & 4 \\ \hline
     5  & 0.43 & 1.04 & 0 & 14 \\ \hline
     6  & 17.90 & 1.56 & 8 & 26 \\ \hline
     7  & 301.26 & 2.57 & 28 & 132 \\ \hline
     8  & 33681.66 & 9.43 & 94 & 320 \\ \hline
     9  &  & 44.97 & 259 & 1224 \\ \hline
     10  &  & 978.67 & 712 & 3594 \\ \hline
  \end{tabular}
\label{tabdatacyclic}
\end{center}
\end{table}

Just as we surpassed our estimates of the
expected number of cone intersections in the generic case, we also
surpassed our estimated ratio in the case of reduced cyclic-$n$.
Table~\ref{conetable} contains experimental results.

\begin{table*}[hbt]
\begin{center}
\caption{Number of cone intersections required for each algorithm,
comparing the definitional algorithm with our pruning algorithm for reduced cyclic-$n$.
The second to last column contains the ratio
predicted by our cost estimate and the final column contains the actual ratio.}
  \begin{tabular}{|r||r|r||r|r|} \hline
    $n$ & Definitional & Pruning & Predicted Ratio & Actual Ratio \\ \hline \hline
     4 & 120  & 44 & 0.25 & 0.36 \\ \hline
     5   &  1850    &  210   &  0.125 & 0.113  \\ \hline 
     6  &  63,981    & 2,040  &  0.0625 & 0.0318 \\ \hline
     7  & 989,751 & 6,272   &  0.03125 & 0.00634 \\ \hline
     8  & 58,155,904 & 39,808  & 0.015625 & 0.000684 \\ \hline
     9  &  & 198,300 & 0.0078125 &  \\ \hline
     10  &   & 1,933,147 & 0.00390625 &  \\ \hline
  \end{tabular}
\label{conetable}
\end{center}
\end{table*}

\subsection{$n$-body and $n$-vortex}

The $n$-body problem~\cite{HJ15} is a classical problem from celestial 
dynamics that states that the acceleration due to Newtonian
gravity can be found by solving a system of equations (\ref{nbodysys}).
These equations can be turned into a polynomial system by clearing the
denominators.
\begin{equation} \label{nbodysys}
\textnormal{\"x}_j = \sum_{i\ne j}\frac{m_i(x_i-x_j)}{r_{ij}^3} ~~~~~~ 1 \le j \le n
\end{equation}

The $n$-vortex problem~\cite{HM09} arose from a generalization of a problem
from fluid dynamics that attempted to model vortex filaments (\ref{nvortexsys}).
Again, these equations can be turned into polynomials through clearing
denominators.

\begin{equation} \label{nvortexsys}
V_i = I\sum_{i\ne j}\frac{\Gamma_j}{z_i-z_j} ~~~~~~ 1 \le j \le n
\end{equation}

Table~\ref{combinedtable} displays experimental results for both the $n$-body
problem and the $n$-vortex problem. We expect to be able to compute
higher $n$ for these benchmark problems when we develop a compiled
version of this code.

\begin{table}[hbt]
\begin{center}
\caption{Experimental results of our new algorithm. Timings are in seconds.
The last column gives the number of cone intersections.}
  \begin{tabular}{|lr||r|r|r|} \hline
   System & $~~~n$ & Pruning Time  &  \#Pretropisms & \#Intersections\\
 \hline \hline
  $n$-body  & 3  &  0.62 & 4 & 121  \\ 
    & 4  &  5.07 & 57 & 25,379  \\ 
    & 5  & 13,111.42  & 2,908 & 18,711,101  \\ \hline 
  $n$-vortex  & 3  &  0.71  & 4 &  87 \\ 
    & 4  &  2.93   & 25 & 10,595  \\ 
    & 5  & 1457.48 & 569 & 5,021,659  \\ \hline
  \end{tabular}
\label{combinedtable}
\end{center}
\end{table}

\section{Conclusion}
To compute all pretropisms of a Laurent polynomial system,
we propose to exploit the connectivity of edge skeletons to prune
the tree of edges of the tuple of Newton polytopes. The horizontal
and vertical pruning concepts we introduce are innovations that reduce the 
computational complexity of the problem.
Our first high level implementation in Sage provides practical evidence
that shows that our new pruning method is better than the definitional
method with a variety of types of polynomial systems.

\bibliographystyle{plain}

\end{document}